\documentclass[preprint,review,12pt]{elsarticle}

\usepackage{graphicx}
\usepackage{url}
\usepackage{amssymb}
\usepackage{lineno}
\usepackage[section]{placeins}

\newcommand{\pt}{\ensuremath{p_{T}}}

\newcommand{\ie}{{\sl i.e.}}

\newcommand{\hbb}{\ensuremath{h\rightarrow b\bar{b}}}

\newcommand{\Pythia}{\textsc{Pythia8}}

\newcommand{\FastJet}{\textsc{FastJet}}
\newcommand{\fjcontrib}{\textsc{FastJet-contrib}}

\author{Riccardo Di Sipio} 
\address{riccardo.disipio@utoronto.ca}

\begin{document}
\begin{frontmatter}

\title{Shower it again, Pythia}

\maketitle

\begin{abstract}
The Parton-Shower algorithm implement in the \Pythia{} generator is applied multiple times to the same parton-level configuration to estimate the stochastic uncertainty affecting large-radius jet substructure variables associated with the stochastic nature of the algorithm. Results are presented in the case of boosted \hbb{} and $t\rightarrow bqq$. The code is publicly available on the repository \url{https://github.com/rdisipio/ReShower.git}
\end {abstract}


\end{frontmatter}


\section*{Introduction}
Among different sources of uncertainty, there are known knowns, things we know we know. There are also known unknowns, some things we know may exist but do not know much about. Notoriously, there are unknown unknowns -- the ones we don't know we don't know. In this paper I want to discuss an example of a unknown knowns, that is to say things we do not single out on purpose, but in fact we know a lot about. The usual argument to ignore them is that they are small compared to other uncertainties or can not be easily factored out. In this work I will focus on one of them, the intrinsic stochastic uncertainty of the parton shower algorithm. What happens if the very same hard-scattering event is showered time and again? Shouldn't we consider the fact that, for each generated event, we observe only a single realization of the process? Such an uncertainty is likely to be covered when a very large sample of events are simulated: one can assume that among millions of events, a number of them lay very close to each other in the phase-space so that those are virtually identical. There may be some outliers for which this is a too strong assumption. Also, this procedure can be used to assign an uncertainty to a single event and may turn out to be useful to train a classifier or other deep learning systems. Finally, it can be viewed as an example of {\sl likelihood-free} inference, in which the probability of an event to be showered following a certain history is conditioned by the parton-level four-momentum of the originating particle.

To restrict the aim of the study to a few concrete examples, a Higgs boson or a top quark are generated always with the same four-momentum, and then decayed by the \Pythia{} Shower Monte Carlo generator \cite{pythia8} in the $b\bar{b}$ or fully-hadronic channel respectively. Afterwards, large-radius jets are reconstructed at hadron level to see how many times the particle is identifiable using jet tagging techniques as a function of kinematic observables such as the jet transverse momentum (\pt) and pseudo-rapidity ($\eta$). One can expect that the decay products of a very high-\pt{} particle will remain collimated in the vast majority of the times, but this picture may change dramatically as the transverse momentum of the original particle approaches the Lorentz boost threshold given by: 
\begin{equation}
    p_{T}^{min} = \frac{2m}{R}
\end{equation}
where $m$ is the mass of the particle and $R$ is the distance parameter of the anti-$k_T$ algorithm used for the jet clustering. In the kinematic region close to the threshold, effects due to the emission of QCD radiation can affect the final state configuration so that a single large-radius jet may be unable to capture all the decay products and hence be ineffective to identify such boosted particles. One of the aims of this paper is to quantify how often this happens in the cases under consideration, and how the resulting event kinematics is affected by the intrinsic stochastic nature of the decay and parton shower processes. 

\section{Pythia Setup}
Higgs bosons and top quarks are individually generated in the {\sl particle-gun} setup that is possible in \Pythia{} 8.240 \cite{pythia8}, \ie{} a single particle is added to the event record with no underlying $pp$ scattering process. Only the hadronic decays \hbb{} and $t\rightarrow bqq$ are allowed. Jets are reconstructed using the anti-$k_{T}$ algorithm \cite{antikt} with distance parameter $R$ = 1.0 implemented in \FastJet{} 3.3.2 \cite{fastjet}. The calculation of the $N$-subjettiness \cite{subjettiness} ratios $\tau_{21}=\tau_2/\tau_1$ and $\tau_{32}=\tau_3/\tau_2$ is implemented in the \FastJet{} plugin distributed with the package \fjcontrib{} 1.041. The $N$-subjettiness is defined as 
\begin{equation}
\tau_N = \frac{1}{d_0}\sum_k p_{T,k} \min \left \{
\Delta R_{1,k}, \Delta R_{2,k}, \Delta R_{N,k}
\right \}
\end{equation}

where the index $k$ runs over the constituent particles in a given jet, $p_{T,k}$ are their transverse momenta, $\Delta R_{J,k} = \sqrt{(\Delta \eta)^2 + (\Delta\phi)^2}$ is the distance in the rapidity-azimuth plane between a
candidate subjet J and a constituent particle $k$, and $d_0$ is a normalization constant defined as $\sum_k p_{T,k} R$, where $R$ is the distance parameter used in the clustering algorithm (in this case $R$=1.0).

\section{Results}
\begin{figure}
\centering
\includegraphics[width=\linewidth]{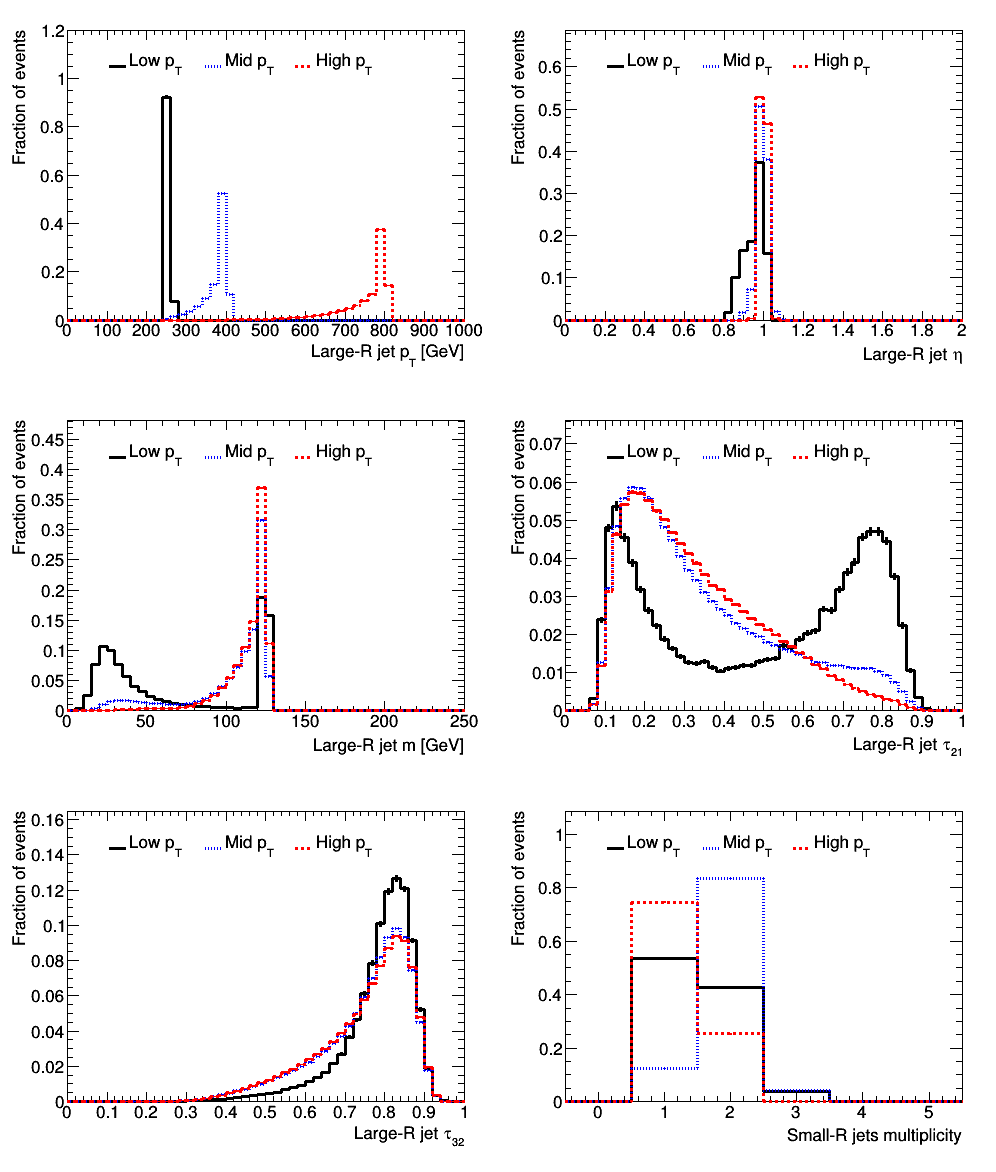}
\caption{\small{The parton-shower algorithm is applied 500,000 times to the same \hbb{} event. The solid black, dotted blue and dashed red lines correspond to \pt{} equal to 250, 400 and 800 GeV respectively.}}
\label{fig:higgs}
\end{figure}

\begin{figure}
\centering
\includegraphics[width=\linewidth]{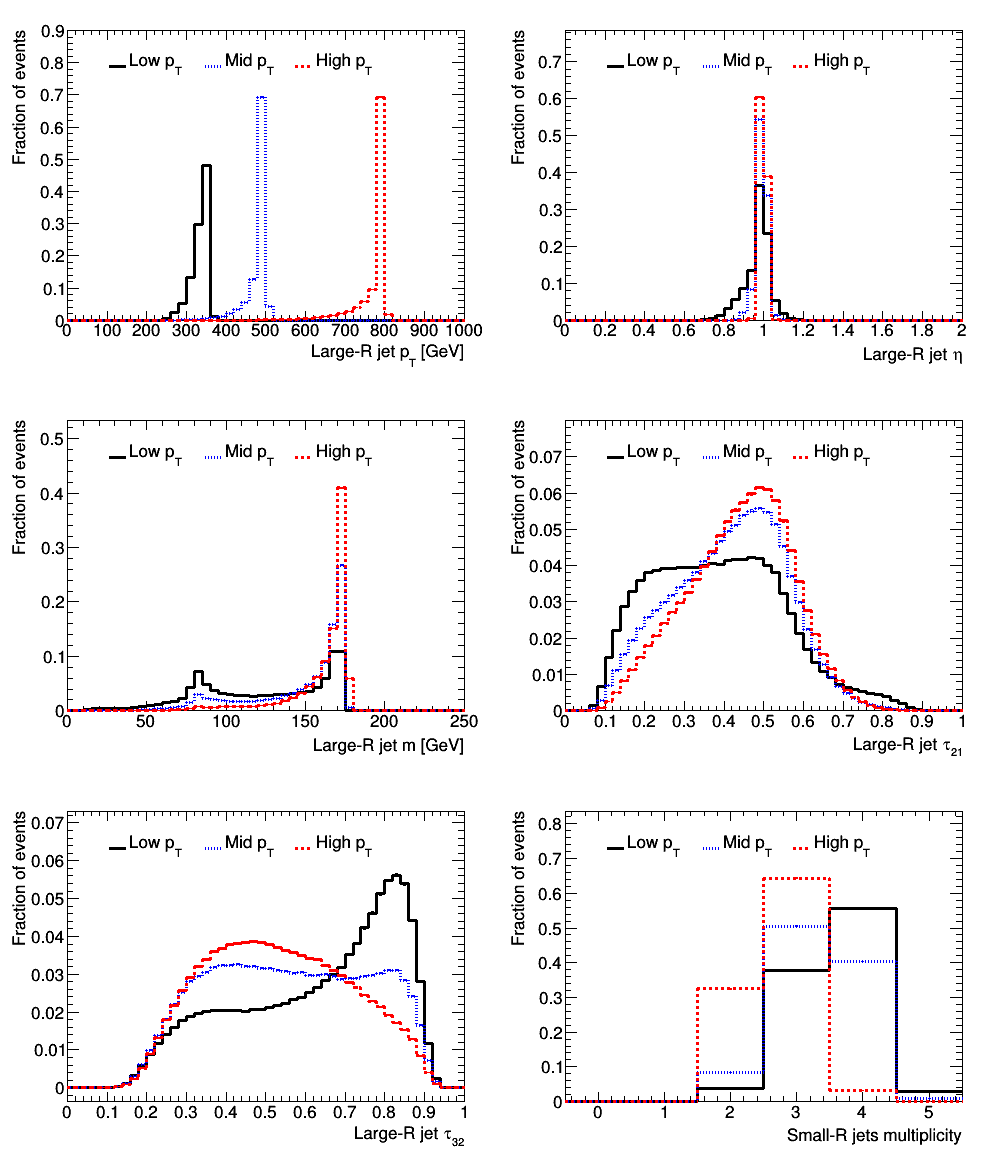}
\caption{\small{The parton-shower algorithm is applied 500,000 times to the same $t\rightarrow bqq$ event. The solid black, dotted blue and dashed red lines correspond to \pt{} equal to 350, 500 and 800 GeV respectively.}}
\label{fig:top}
\end{figure}

The effect of the repeated application of the parton-shower algorithm is presented in Figures \ref{fig:higgs} and \ref{fig:top}. Out of many possibilities, six observables have been identified to summarize concisely the jet kinematics: transverse momentum ($p_T$), pseudorapidity ($\eta$), mass ($m$), $N$-subjettiness ratios $\tau_{21}$ and $\tau_{32}$, and the number of geometrically-matched sub-jets, \ie{} jets clustered with the anti-$k_T$ algorithm with $R$=0.4, whose axis is closer than $\Delta$ R = 1.0 from the large-$R$ jet axis.    

Figure \ref{fig:higgs} shows the result for a boosted \hbb{} event for a transverse momentum of the Higgs boson equal to 250, 400 and 800 GeV and pseudorapidity $\eta$ = 1.0. The parton shower algorithm is applied 500,000 times to each event. The smearing due to QCD radiation has a mild effect for low Higgs boson $p_T$, but presents a more pronounced and asymmetric tail as the transverse momentum increases. The effect can be quantified in terms of the width of the distribution of about 100 GeV at $p_T$ = 400 GeV, and about twice as large for $p_T$ = 800 GeV.
The most dramatic effects can be seen in the mass and substructure variables distributions. It is evident that low-\pt{} Higgs bosons are ''un-boosted'' about 60\% of the times and the resulting leading large-$R$ jet does not capture the full decay. Interestingly, the number of small-radius ($R=0.4$) jets peaks at 2 for a mid-\pt, but at one for a very high transverse momentum, a telltale sign of the Lorentz boost. The distribution of $\tau_{21}$ shows that a cut applied at $\tau_{21}<0.5$ is effective to identify two-prong decays only when the \pt{} of the Higgs is significantly higher than the Lorentz threshold.

Figure \ref{fig:top} shows similar results in the case of a $t\rightarrow Wb \rightarrow qqb$ decay with possible values of the top \pt{} equal to 350, 500 and 800 GeV and pseudorapidity $\eta$ = 1.0. In the case of low top transverse momentum, a secondary peak in the mass distribution corresponding to a boosted $W$ boson is clearly visible, while at high-\pt{} the large-$R$ jet is able to capture all the decay products. Also, the number of small-radius jets peaks at different values as a function of the top $p_T$. The substructure variable $\tau_{32}$ tends to be more symmetric only in the high-\pt{} regime. The substructure variables indicate that at low-\pt{} about 60\% of the events do not look like a three-prong decay and would be rejected by a cut $\tau_{32} < 0.6 $.

\begin{figure}
\centering
\includegraphics[width=\linewidth]{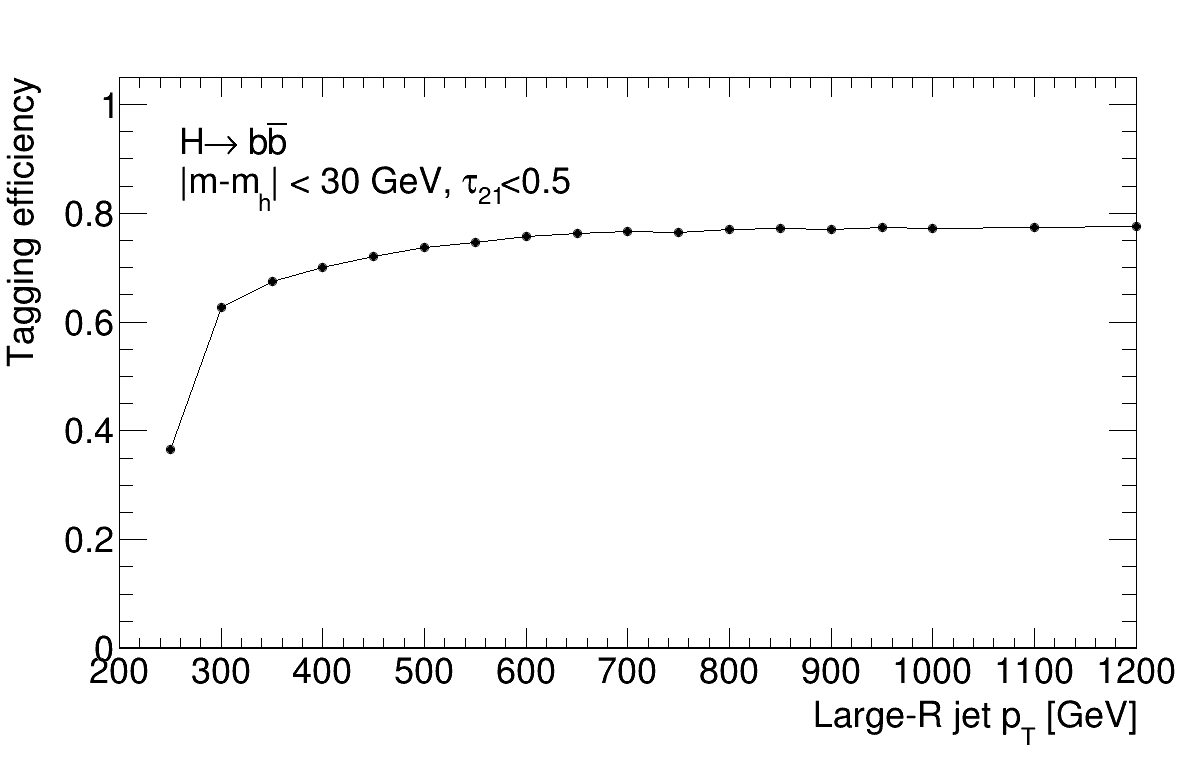}
\caption{\small{Higgs-tagging efficiency as a function of the parton transverse momentum.}}
\label{fig:eff:higgs}
\end{figure}

\begin{figure}
\centering
\includegraphics[width=\linewidth]{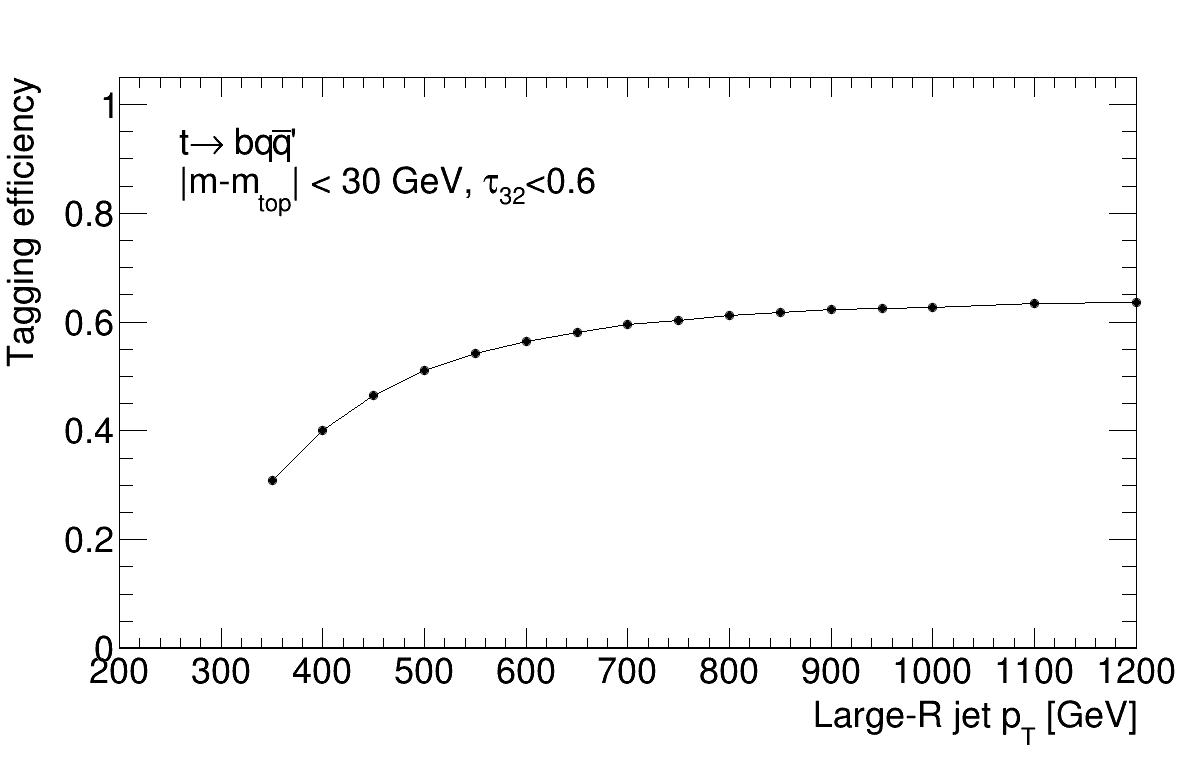}
\caption{\small{Top-tagging efficiency as a function of the parton transverse momentum. Red lines indicate the region selected by the simple tagger.}}
\label{fig:eff:top}
\end{figure}

\begin{figure}
\centering
\includegraphics[width=0.5\linewidth]{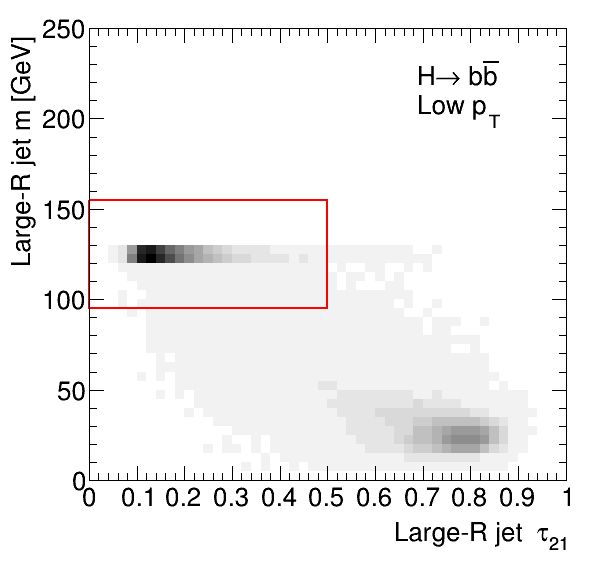}
\includegraphics[width=0.5\linewidth]{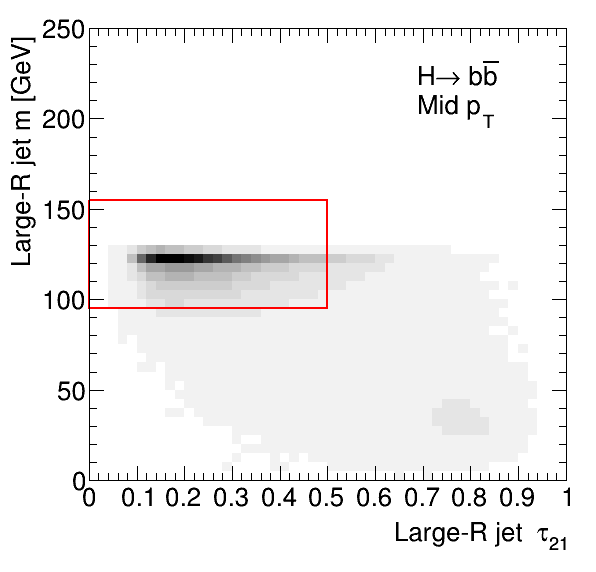}
\includegraphics[width=0.5\linewidth]{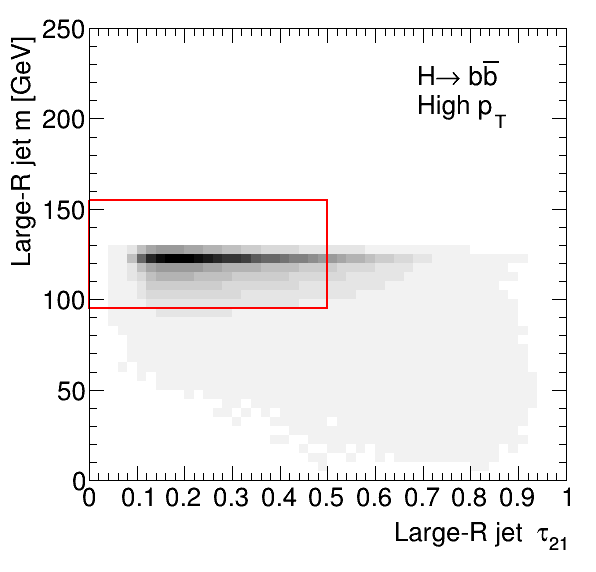}
\caption{\small{Correlations between the jet mass and substructure variable $\tau_{21}$ for \hbb{} events for different values of the Higgs boson transverse momentum. Red lines indicate the region selected by the simple tagger.}}
\label{fig:2d:higgs}
\end{figure}

\begin{figure}
\centering
\includegraphics[width=0.5\linewidth]{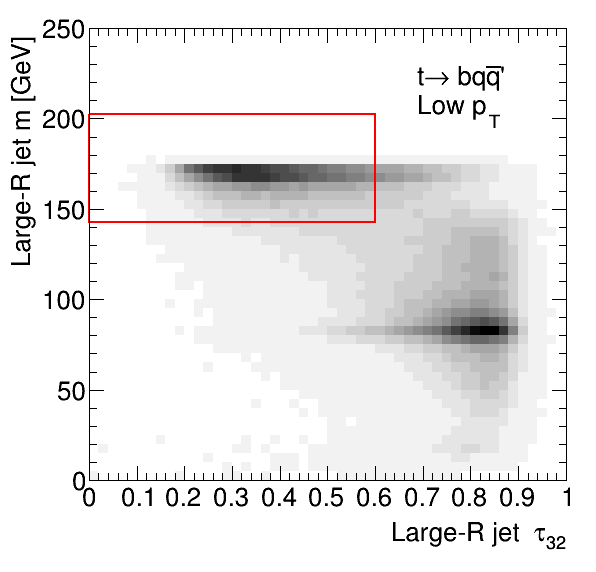}
\includegraphics[width=0.5\linewidth]{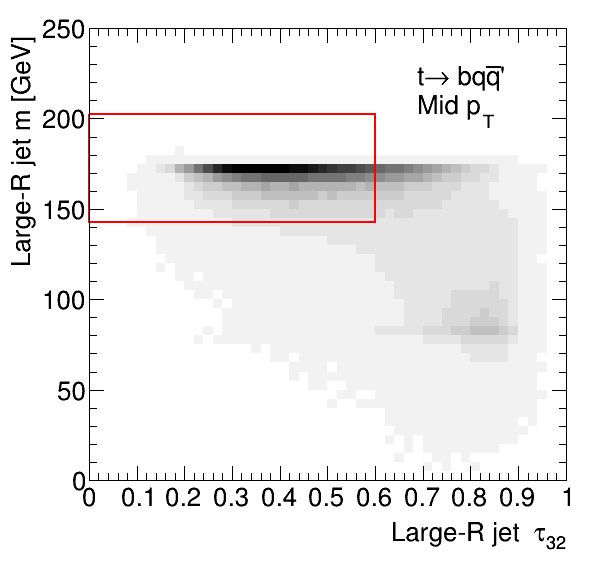}
\includegraphics[width=0.5\linewidth]{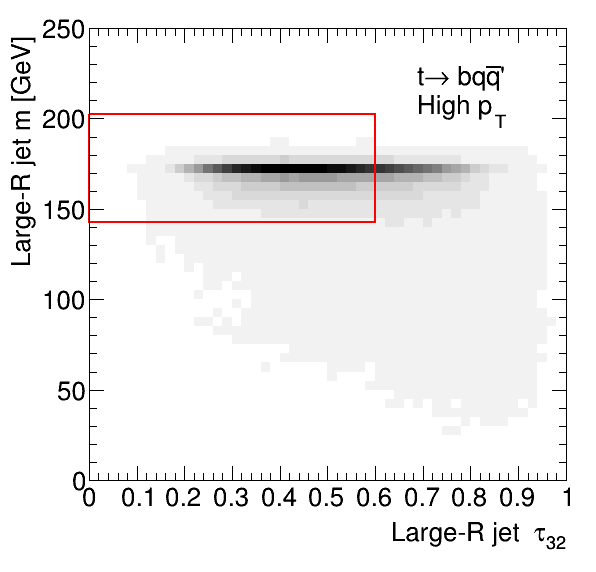}
\caption{\small{Correlations between the jet mass and substructure variable $\tau_{32}$ for $t\rightarrow bqq$ events for different values of the top quark transverse momentum. Red lines indicate the region selected by the simple tagger.}}
\label{fig:2d:top}
\end{figure}

Figures \ref{fig:eff:top} and \ref{fig:eff:higgs} show respectively the top- and Higgs-tagging efficiency as a function of the parton transverse momentum. A simple tagger algorithm is used to demonstrate the effect, defined as: 
\begin{itemize}
\item $|m_{jet}-m_{h,PDG}|<$ 30 GeV and $\tau_{21}<$ 0.5 for the Higgs boson.
\item $|m_{jet}-m_{t,PDG}|<$ 30 GeV and $\tau_{32}<$ 0.6 for the top quark. 
\end{itemize}
Neither is fully efficient, but both reach a plateau in the mid-to-high-\pt{} region. 
Finally, figures \ref{fig:2d:higgs} and \ref{fig:2d:top} show the correlation between the jet mass and the substructure variable used for tagging for the Higgs boson and the top quark respectively. As a result, the tagger selects the region where the decay products are contained in the large-$R$ jet, but in both cases at low transverse momentum about half of the events fall outside that kinematic region, resulting in a jet with lower mass and high subjettines ratio. In the case of the top quark, the jet mass around $80$ GeV indicates that such jet corresponds to the boosted $W$ boson. 

\section{Conclusions}
The effect of the intrinsic stochastic nature of the parton shower algorithm was studied in the case of Higgs bosons and top quarks with high transverse momentum. The uncertainty associated to this effect is usually considered to be small when dealing with large samples of simulated events and traditionally is neglected. However, it is shown that simply due to the intrinsic stochastic nature of the parton shower process, top quarks and Higgs bosons with low transverse momentum may not be reconstructed using a single large-$R$ jets up to 60\% of the cases.
On top of what is presented in the examples above, changing the parameters of the showering algorithm would also affect the distributions obtained by repeatedly applying the parton shower to the same events. Particles with a transverse momentum close to the Lorentz boost threshold are subject to large effects due to the QCD radiation and the kinematics of their decay product can appear largely different to the point that the traditional approach to identify boosted resonances can be completely ineffective. 
The method introduced in this paper can potentially lead to a novel approach to quantify the statistical uncertainty due to parton shower and other similar algorithms that can not be easily described by a likelihood function, with particular interest in border cases and the tuning of boosted-object taggers.

\section*{Acknoledgments}
I would like to thank Pekka Sinervo, Kyle Cormier and Francesco Span\`{o} for the very useful discussions about this topic. We'll always have Monte Carlo. I acknowledge the support of the Natural Sciences and Engineering Research Council of Canada (NSERC).

\end{document}